\documentclass[final,10pt,1p]{elsarticle}

\usepackage{amssymb}
\usepackage[T1]{fontenc} % Use 8-bit encoding that has 256 glyphs
\usepackage{fourier} % Use the Adobe Utopia font for the document - comment this line to return to the LaTeX default
\usepackage[english]{babel} % English language/hyphenation
\usepackage{amsmath,amsfonts,amsthm} % Math packages
\usepackage{appendix}

\usepackage{amssymb}
\usepackage[all]{xy}
\usepackage{enumerate}
\usepackage{tikz}
\usepackage{mathrsfs}
\usepackage[english]{babel}
\usepackage{multirow}
\usepackage{float}
\usepackage{enumitem}

\newtheorem{innercustomthm}{Theorem}
\newenvironment{customthm}[1]
  {\renewcommand\theinnercustomthm{#1}\innercustomthm}
  {\endinnercustomthm}

\newtheorem{innercustomlemma}{Lemma}
\newenvironment{customlemma}[1]
  {\renewcommand\theinnercustomlemma{#1}\innercustomlemma}
  {\endinnercustomlemma}

\newtheorem{innercustomcorollary}{Corollary}
\newenvironment{customcorollary}[1]
  {\renewcommand\theinnercustomcorollary{#1}\innercustomcorollary}
  {\endinnercustomcorollary}

\newtheorem{innercustomdef}{Definition}
\newenvironment{customdef}[1]
  {\renewcommand\theinnercustomdef{#1}\innercustomdef}
  {\endinnercustomdef}

\newtheorem{innercustomprop}{Proposition}
\newenvironment{customprop}[1]
  {\renewcommand\theinnercustomprop{#1}\innercustomprop}
  {\endinnercustomprop}

\journal{Physica D}

\begin{document}

\begin{frontmatter}

%% Title, authors and addresses

%% use the tnoteref command within \title for footnotes;
%% use the tnotetext command for theassociated footnote;
%% use the fnref command within \author or \address for footnotes;
%% use the fntext command for theassociated footnote;
%% use the corref command within \author for corresponding author footnotes;
%% use the cortext command for theassociated footnote;
%% use the ead command for the email address,
%% and the form \ead[url] for the home page:
%% \title{Title\tnoteref{label1}}
%% \tnotetext[label1]{}
%% \author{Name\corref{cor1}\fnref{label2}}
%% \ead{email address}
%% \ead[url]{home page}
%% \fntext[label2]{}
%% \cortext[cor1]{}
%% \address{Address\fnref{label3}}

\title{Synchronization of finite-state pulse-coupled oscillators}

%% use optional labels to link authors explicitly to addresses:
%% \author[label1,label2]{}
%% \address[label1]{}
%% \address[label2]{}

\author[1]{Hanbaek Lyu }
\address{Department of Mathematics, The Ohio State University, Columbus, OH 43210 (yu.1242@osu.edu)}

\newtheorem{problem}{Problem}
\newtheorem{proposition}{Proposition}[section]
\newtheorem{lemma}{Lemma}
\newtheorem{corollary}{Corollary}
\newtheorem{definition}{Definition}[section]
\newtheorem{theorem}{Theorem}
\newtheorem{remark}{Remark}[section]
\newtheorem{ex}{Example}
\newtheorem{exercise}{Exercise}
\newtheorem{note}{Note}

\def\bibsection{\section*{References}}

\begin{abstract}
We propose a novel generalized cellular automaton(GCA) model for discrete-time pulse-coupled  oscillators and study the emergence of synchrony. Given a finite simple graph and an integer $n\ge 3$, each vertex is an identical oscillator of period $n$ with the following weak coupling along the edges: each oscillator inhibits its phase update if it has at least one neighboring oscillator at a particular "blinking" state and if its state is ahead of this blinking state. We obtain conditions on initial configurations and on network topologies for which states of all vertices eventually synchronize. We show that our GCA model synchronizes arbitrary initial configurations on paths, trees, and with random perturbation, any connected graph. In particular, our main result is the following local-global principle for tree networks: for $n\in \{3,4,5,6\}$, any $n$-periodic network on a tree synchronizes arbitrary initial configuration if and only if the maximum degree of the tree is less than the period $n$. 
\end{abstract}

\begin{keyword}
%% keywords here, in the form: keyword \sep keyword

%% PACS codes here, in the form: \PACS code \sep code

%% MSC codes here, in the form: \MSC code \sep code
%% or \MSC[2008] code \sep code (2000 is the default)
Synchronization \sep  pulse-coupled oscillators \sep  generalized cellular automata \sep digital clock synchronization\sep self-stabilization \sep path \sep  tree \sep absorbing chain 

\end{keyword}

\end{frontmatter}

%% \linenumbers

%% main text
\section{Introduction}
\label{intro}

The emergence of collective behavior from locally interacting many-agent systems is a pervasive phenomenon in nature and has raised strong scientific interests \cite{strogatz2001exploring}. In particular, synchronization in systems of pulse-coupled oscillators(PCOs) is a fundamental issue in physics, biology, and engineering. Examples from nature include synchronization of fireflies \cite{buck1938synchronous}, neurons in the brain \cite{tateno2007phase}, and the circadian pacemaker cells \cite{enright1980temporal}). These are all examples of complex systems, which achieve desired system behavior by an aggregation of decentralized interactions between individual agents. This "bottom-up" approach became popular as a paradigm for studying complex systems after the celebrated Boids model by Reynolds \cite{reynolds1987flocks}, which beautifully demonstrated the flocking of birds or fishes in such a framework. Nowadays the technique of decentralized control finds its use in cooperative control of networked dynamical systems, from robotic vehicle networks to electric power networks to synthetic biological networks (\cite{tanner2003stable}, \cite{antsaklis2004guest}, \cite{antsaklis2007special}, \cite{bullo2009special}, \cite{mesbahi2010graph}, \cite{nair2007stable}).

\qquad A discrete-time deterministic dynamical system on a network of finite-state machines with a locally defined homogeneous transition map is called a generalized cellular automaton(GCA), commonly known as a cellular automaton(CA) when the network topology is taken to be a lattice \cite{wolfram1984cellular}. GCAs can exhibit striking spatio-temporal patterns in spite of their simplicity \cite{langton1990computation}, and are gaining growing interest as an alternative paradigm for modeling complex systems \cite{chopard1998cellular}, \cite{haefner2005modeling}. Some extensively studied models include lattice gas automaton for simulating fluid flows \cite{wolf2000lattice}, Greenburg-Hastings model for excitable medium \cite{fisch1993metastability}, and Griffeath's cyclic cellular automaton, which shows clustering on $\mathbb{Z}$ and autowave behavior on $\mathbb{Z}^{2}$ \cite{fisch1990cyclic}. To the author's knowledge, however, there has not been a direct attempt to use GCAs to study the synchronization of PCOs. 

\qquad We can roughly classify the literature on coupled non-linear oscillators by the following four parameters: time, space, coupling, and dynamics. First, most classical studies on this subject use ODE models, which have continuous-time, continuous-space, deterministic dynamics, and phase-coupling. A pioneering work was done by Winfree \cite{winfree1967biological}, and then Kuramoto model has become a paradigm in this area \cite{acebron2005kuramoto}, \cite{strogatz2000kuramoto}. In case of biological oscillators, one usually assumes the mutual coupling is episodic and pulselike (\cite{ermentrout1990oscillator}, \cite{glass1979simple}, \cite{pavlidis2012biological}, \cite{winfree2001geometry}). Peskin \cite{peskin1975mathematical} studied a system of pulse-coupled oscillators(PCOs) to model cardiac pacemaker cells, and later Mirollo and Strogatz \cite{mirollo1990synchronization} generalized his model and showed that synchronization is guaranteed for almost all initial configurations when the oscillators are all-to-all connected. A recent work of Nishmura and Friedman \cite{nishimura2011robust} further generalizes their model, and derives a condition on the initial configuration to guarantee synchronization for arbitrary connected topology. To obtain synchronization for both arbitrary connected topology and initial configuration, Klingmayr et. al \cite{klinglmayr2012guaranteeing} studied a discrete-time continuous-space system of PCOs with stochastic dynamics. DeVille and Peskin \cite{deville2008synchrony} studied all-to-all networks of a discrete and stochastic version of Peskin's excitatory model in \cite{peskin1975mathematical}.

\qquad In computer science, a challenge in discrete deterministic coupled oscillators is known as the\textit{digital clock synchronization problem}, which is to devise a protocol to achieve synchronization on a network of digital clocks that are synchronously updated. Two properties are highly desirable for such protocols: 1) it uses constant number of states on each local clock to ensure scalability, and 2) it synchronizes arbitrary configuration for initial synchronization and fault-tolerance. The second property is called self-stabilization, first proposed by Dijkstra \cite{dijkstra1982self}, which is equivalent to requiring that the set of desirable system states for a protocol is a global attractor in the corresponding discrete dynamical system. Such protocols are readily applicable in wireless sensor networks, for example \cite{sundararaman2005clock}. Nevertheless, it is well-known that there is no such protocol with both properties 1) and 2) that works on arbitrary connected networks \cite{dolev2000self}. While there does exist a protocol with property 2) for arbitrary network topology but with a bounded number of states that depends on the network \cite{arora1992maintaining}, the more relevant work is done by Herman and Ghosh \cite{herman1995stabilizing}; they presented a 3-state protocol on trees, which can be regarded as a 3-state GCA model for \textit{phase}-coupled oscillators.  

\qquad In this paper, we propose GCA models for a discrete system of PCOs and study their network behavior. We call our models the \textit{firefly networks}, due to our initial motivation to understand the emergence of synchronous blinking of fireflies. These are discretized versions of previously studied models mentioned earlier (in particular, the continuous model of Nishimura and Friedman \cite{nishimura2011robust}) and enjoy some of the similar behavior(in particular, Lemma \ref{widthlemma}). Moreover, our model can be regarded as a protocol for digital clock-synchronization that uses constant number of states. Our main results tell us that for some classes of network topologies synchrony is guaranteed to emerge, but there are also examples of connected networks where synchrony may fail to emerge. These results contrast with the models that incorporate a certain type of stochasticity mentioned earlier, for which emergence of synchrony with probability 1 is derived for all connected network of finitely many oscillators \cite{klinglmayr2012guaranteeing}. We also obtain universal synchrony for a randomized version of our model as a corollary of our deterministic results. 
 
\qquad In the rest of the introduction, we give a definition of our GCA model together with some illustrating examples and our main results. Some of the results are derived for the firefly system defined here, and some for more general types of GCA. 
\begin{customdef}{1}
	Let $G=(V,E)$ be a finite simple graph and fix $n\ge 3$. Let $\mathbb{Z}_{n}=\mathbb{Z}/n\mathbb{Z}$ with linear ordering $0<1<2<\cdots<n-1$. An $n$-\textit{configuration} is a map $X:V\rightarrow \mathbb{Z}_{n}$. Let $b(n)=\lfloor \frac{n-1}{2} \rfloor$ be the blinking state. The time evolution of a given initial configuration is given by the \textit{firefly transition map} $\tau_{G}:X\mapsto X'$, defined as follows :
\begin{equation}
	X'(v)=\begin{cases}
X(v) & \text{if $X(v) > b(n)$ and $v$ is adjacent to some vertex of state $b(n)$}\\
X(v)+1 & \text{otherwise}\end{cases}.
\end{equation}
The discrete-time dynamical system on $G$ generated by the iteration of this transition map is called the \textit{$n$-periodic firefly network on $G$}. The firefly network on $G$ starting from an \textit{initial configutation} $X_{0}$ is denoted $(G,\tau_{G},X_{0})$ or $(G,X_{0})$ in short, and the configuration after $t\in \mathbb{N}$ iterations of the transition map will be denoted $X_{t}$. We call the unit of time "second". The sequence $(X_{t})_{t=0}^{\infty}$ will be called the \textit{orbit of $X_{0}$ on $G$}. We say \textit{$(G,X_{0})$ synchronizes} or \textit{$X_{0}$ synchronizes} if there is $N\in \mathbb{N}$ such that $X_{t}$ is a constant function for all $t\ge N$. We say $G$ is \textit{$n$-synchronizing} if every $n$-configuration on $G$ synchronizes. 
\end{customdef}

\qquad In words, our transition rule can be interpreted as follows: in a network of $n$-state identical oscillators, each oscillator updates from state $s$ to $s+1(\text{mod $n$})$ unless it senses a blinking state and notices that its phase is ahead of the blinking neighbor, in which case it waits for 1 second without update. In this case, we say that the vertex is \textit{pulled} by its blinking neighbor.

\begin{figure*}[h]
	\centering
	\includegraphics[width=0.95 \linewidth]{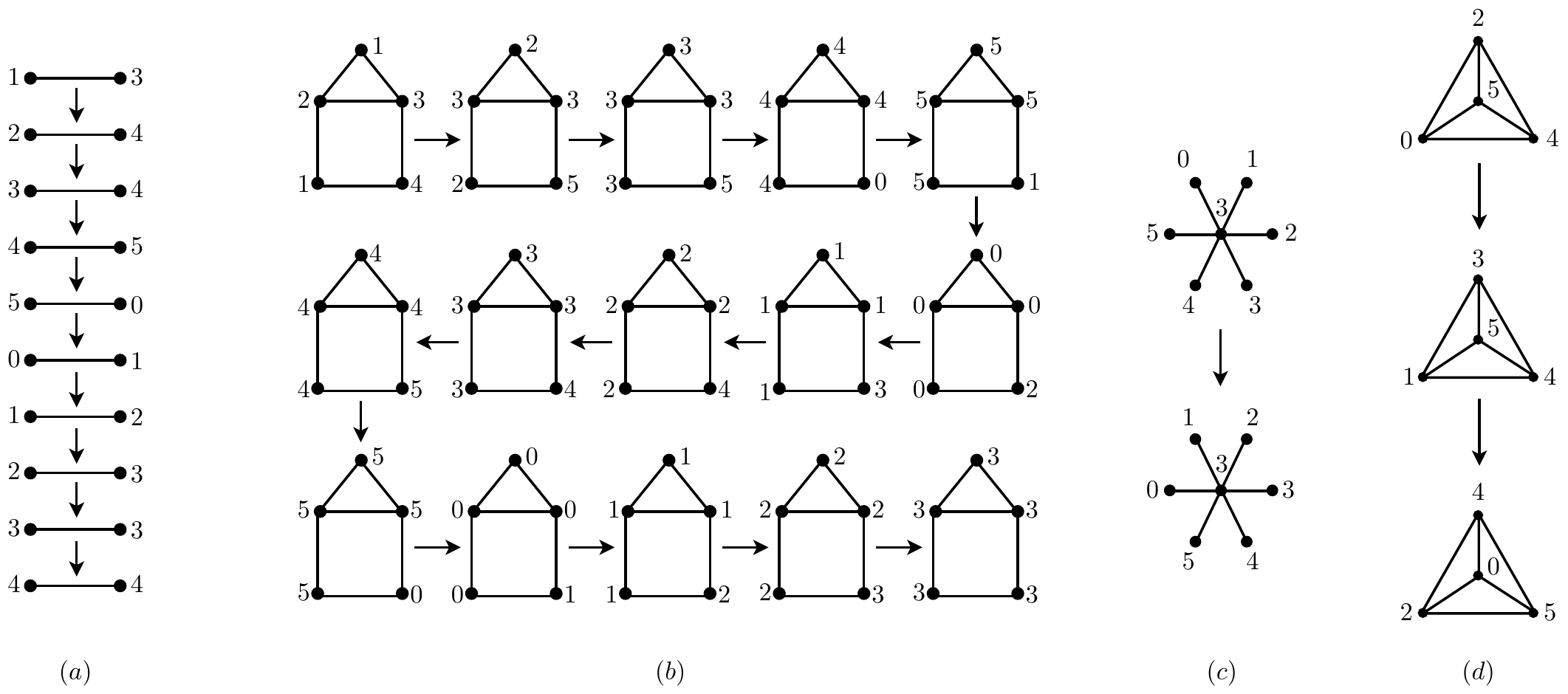}
	\caption{Two examples of synchronizing 6-periodic networks are shown in (a) and (b). In (c) and (d), the last configurations are symmetric to the initial ones, so the networks do not synchronize. }
	\label{ex1}
\end{figure*}

\begin{figure*}[h]
	\centering
	\includegraphics[width=8cm,height=2cm]{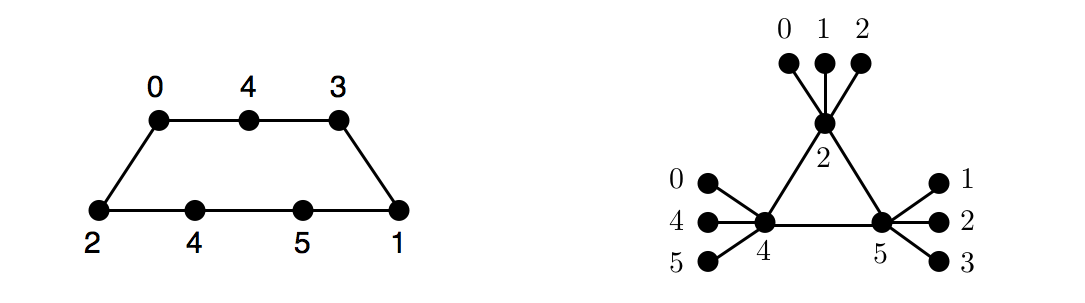}
	\caption{More examples of non-synchronizing 6-periodic firefly networks.}
	\label{nonsync}
\end{figure*}

\qquad Our network model is a discrete-time dynamical system with finite configuration space. Hence, every trajectory must converge to a limit cycle. Limit cycles can be either a synchronous or asynchronous periodic orbit, as illustrated in the examples of 6-periodic networks in Figures \ref{ex1} and \ref{nonsync}. Note that $b(6)=2$ is the blinking state in this case, so every vertex of state $3,4,$ or $5$ with a state 2 neighbor stops evolving for 1 second and all the other vertices evolves to the next state. 

\qquad As illustrated in the examples in Figure \ref{ex1} and \ref{nonsync}, whether a network $(G,X_{0})$ synchronizes depends on the structure of $G$, initial configuration $X_{0}$,  and also on the period $n$ as we will see at the end of Section 6. We find conditions on $G,n$, and $X_{0}$ to guarantee synchronization. In particular, we find conditions on $G$ and $n$ such that our GCA model for synchronization is self-stabilizing, i.e., it synchronizes arbitrary $n$-configurations on $G$. It appears that paths are inherent to our model, in the sense of the following theorem: 

\begin{customthm}{2}\label{path}
	Every path $P$ is $n$-synchronizing for all $n\ge 3$. Furthermore, the maximum synchronization time, i.e., the largest possible number of steps to synchronize an initial configuration is linear in the size of $P$. More precisely, let $P$ be a path of $m$ vertices, and let $T_{P}(m)$ be the maximum synchronization time. Then we have the following linear bound 
	\begin{equation}
	 n\left( \frac{n}{2}-1+m \right) \le T_{P}(m)\le (m-1)\left( \frac{n^{2}}{2}+2n-2\right).
	\end{equation}
\end{customthm}

\qquad As one tries to expand the class of $n$-synchronizing graphs, the class of finite trees would be a natural choice that includes finite paths. However, as illustrated by example (c) in Figure \ref{ex1}, there exists a tree with a bad 6-configuration, which never synchronizes. In this counterexample, the obvious obstacle to obtain synchrony on trees is that a vertex with many neighbors could stop updating when it is constantly pulled by its neighbors. Namely, let $v$ be a vertex of a finite tree $T$ with degree $\ge n$, and let $T_{1},\cdots,T_{m}$ be the connected components of $T-v$, the graph with $v$ and its edges are removed form $T$. Note that $m\ge n$. Assign state $i(\text{mod $n$} )$ to every vertex of $T_{i}$, and assign any state $>n/2$ to vertex $v$. Then $v$ never blinks and each component $T_{i}$ never get pulled by $v$, which is essentially the counterexample in Figure \ref{ex1} (c). Hence if every $n$-configuration on $T$ synchronizes, it is necessary that $T$ has maximum degree $<n$. 

\qquad A non-trivial result is the converse, namely, the "if" part of the following theorem

\begin{customthm}{3}\label{treedegree}
	Let $T$ be a tree and let $n\in \{3,4,5,6\}$. Then $T$ is $n$-synchronizing if and only if the maximum degree of $T$ is strictly less than $n$. 
\end{customthm}

This result gives a necessary and sufficient local condition on network topology that guarantees global synchrony. Observe that the key feature of the counterexample Figure \ref{ex} (c) is that some vertex could stop blinking eventually due to constant pullings from its neighbors, and in turn those neighbors stop being pulled by this vertex. This isolates the connected components of $T-v$ due to the tree structure. Indeed, the following theorem shows that for certain periods, this is a crucial factor to guarantee synchrony on trees. 

\begin{customthm}{4}\label{blinkingtree}
	Let $T$ be a tree and let $X_{0}$ be a $n$-configuration for some $n\in\{3,4,5,6\}$. Then $(T,X_{0})$ synchronizes if and only if every vertex of $T$ blinks infinitely often in the dynamic. 
\end{customthm}

An easy way to achieve such blinking property is to make the maximum degree of the underlying graph to be less than the period $n$, as in the following lemma. 

\begin{customlemma}{5}\label{deglemma}
	Let $G=(V,E)$ be a graph and let $u$ be a vertex. Suppose $\deg_{G}(u)< n$. Let $X_{0}$ be any $n$-configuration on $G$. Then $u$ blinks infinitely often in the dynamic $(G,X_{0})$. 
\end{customlemma}

Now Theorem \ref{treedegree} follows easily from Theorem \ref{blinkingtree} and Lemma \ref{deglemma}; for if $n\in\{3,4,5,6\}$ and $T$ is a tree with maximum degree $<n$, then for any $n$-configuration $X_{0}$ every vertex blinks infinitely often in the dynamic, and hence the configuration synchronizes by the theorem. 

\begin{figure*}[h]
    \centering
          \includegraphics[width=0.5 \linewidth]{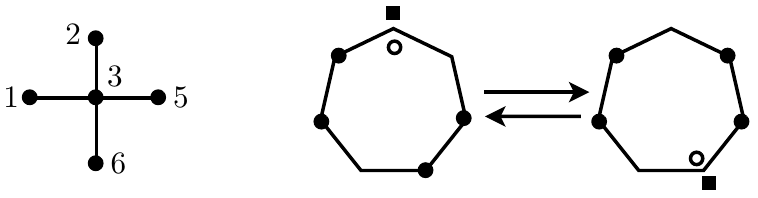}
          \caption{A counterexample for the blinking tree theorem for $n=7$. Diagrams in the right represents the dynamic on this counterexample in an alternative representation called relative circular representation, which will be introduced in Definition 2.3. The square and dots here denote the relative state of the center and the leaves, respectively. }
          \label{n7counterex}
\end{figure*}

\qquad Our proof of Theorem \ref{blinkingtree} when $n=6$ is long and technical, so it will be omitted in this paper. It should be mentioned that the local-global principle of Theorem 2 fails to hold for $n=7$. Consider a star of degree 4 with following 7-periodic initial configuration in Figure \ref{n7counterex}. Every vertex blinks infinitely often in the network but it does not synchronizes. 
Verifying Theorem \ref{blinkingtree} for $n\ge 8$ is still open. 

\qquad The above results tell us that for some classes of network topologies synchrony is guaranteed to emerge, possibly depending on the period $n$, whereas there are also some connected networks where synchrony may fail to emerge. While it is interesting but difficult to characterize the class of synchronizing graphs, if we introduce a certain type of stochasticity in our deterministic model, we can easily show that the network on any connected graph synchronizes with probability 1. The intuition behind this randomization is that by randomness we can disregard subtle influence of the underlying topology on the global dynamic and break the symmetry of non-synchronizing orbits. The following result illustrates this observation; assuming stochastic reception of signals in our deterministic model, we quickly obtain the following result as an application of Theorem \ref{path}.

\begin{customthm}{6}\label{stochastic}
	Let $G$ be any finite connected graph and fix $n\ge 3$. Let $X_{0}$ be any $n$-configuration on $G$. Suppose that each edge $e$ in $G$ is present at each instant independently with a fixed probability $p_{e}\in (0,1)$. Then the dynamic $(G,X_{0})$ synchronizes with probability 1. 
\end{customthm}

\qquad In section 2 we discuss a more general version of the GCA of Definition 1, give a geometric representation of the dynamic, and discuss a fundamental observation on stable manifolds on which synchronization is guaranteed for arbitrary connected topology. And then we mention a characteristic property of our model as an inhibitory system, which enables inductive arguments in the later sections. In section 3, we establish a key lemma(Lemma \ref{branchwidthlemma}) in this paper and prove Theorem \ref{path}. In the following section, Section 4, we discuss transient and recurrent local configurations based on the concept of Poincar\'e return map, and prove Theorem \ref{treedegree} for $n\in \{3,4,5\}$. In section 5, we randomize our deterministic model and show that the resulting Markov chain is in fact an absorbing chain with synchrony being the unique absorbing state. Hence such randomized versions of our model synchronizes with probability 1 on any connected graphs.

\qquad In the following discussions, we assume every graph is finite, simple, and connected, unless otherwise mentioned. If $S,H$ are vertex-disjoint subgraphs of $G$, then $S+H$ is defined by the subgraph obtained from $S\cup H$ by adding all the edges in $G$ between $S$ and $H$. On the other hand, $S-H:=S-V(H)$ denotes the subgraph of $G$ obtained from $S$ by deleting all vertices of $H$ and edges incident to them. If $v\in V(H)$ and $H$ is a subgraph of $G$, then $N_{H}(v)$ denotes the set of all neighbors of $v$ in $H$ and $\deg_{H}(v):=|N_{H}(v)|$.

\section{Generalities, relative circular representation, and the width lemma}

\qquad We begin our discussion with a general consideration on GCA models for finite-state coupled oscillators. Let $G=(V,E)$ be a graph and fix $n\ge 3$, and let $S(G,n)=\mathbb{Z}_{n}^{|V|}$ be the set of all $n$-configurations on $G$. A transition map $\tau_{G}$ on $S(G,n)$ then defines deterministic dynamics on $G$. Let $\tau_{G}$ be the firefly transition map given in Definition 1. It has the following natural properties that would be required for any GCA model for coupled oscillators:

\begin{description}[noitemsep]
	\item{(i)} (\textit{isolation}) If $v\in V$ is isolated in $G$, then $\tau_{G}(X)(v)=X(v)+1$;
	\item{(ii)} (\textit{local dependence}) If $v\in V$ and $e\in E$ is any edge that is not incident to $v$, then $\tau_{G}(X)(v)=\tau_{G-e}(X)(v)$. 
\end{description}

Moreover, the coupling given by $\tau_{G}$ is pulse-coupling by which we mean the following property:
\begin{description}[noitemsep]
	\item{(iii)} (\textit{pulse-coupling}) There exists a unique state $b(n)\in \mathbb{Z}_{n}$, called the \textit{blinking state}, such that for each $v\in V$, we have $\tau_{G}(X)(v)=X(v)+1$ if no neighbor of $v$ has state $b(n)$. 
\end{description}

In words, an oscillator evolves to the next state if there is no blinking neighbor, as a non-blinking firefly will not affect its neighboring fireflies or a neuron will not be affected by its non-spiking neighboring neurons. A \textit{discrete system of $n$-periodic PCO(pulse-coupled oscillator)s} is an assignment $\tau:G\mapsto \tau_{G}$ for each finite connected graph $G$ to a transition map $\tau_{G}$ on the set of $n$-configurations on $G$ satisfying conditions (i)-(iii). Hence the $n$-periodic firefly network defined in Definition 1 is a discrete system of PCOs with $b(n)=\lfloor \frac{n-1}{2} \rfloor$ being the blinking state.

\qquad Note that since the state space $\mathbb{Z}_{n}^{V}$ is finite and the dynamic is deterministic, for any system of oscillators on a graph $G=(V,E)$ with any initial configuration, the trajectory must converge to a periodic orbit. The simplest possible and desired orbit is synchrony, where all oscillators have identical states. We wish to achieve global synchrony by an aggregation of local efforts to obtain mutual synchrony. Hence it is natural to require that the system $\tau$ synchronizes two coupled oscillators regardless of initial configuration. Our model indeed has this property with local monotonicity in the sense of property (iv) below:

\begin{customdef}{2.1}
	Let $G=(V,E)$ be a graph, and let $X:V\rightarrow \mathbb{Z}_{n}$ be an $n$-configuration. Let $u,v$ be two vertices in $G$. The \textit{clockwise displacement of $v$ from $u$ in configuration $Y$} is defined by 
\begin{equation*}
	\delta_{X}(u,v):= X(v)-X(u) \,\,(\text{mod $n$}). 
\end{equation*}
If $X_{0}$ is an initial configuration, we write $\delta_{t}(u,v):=\delta_{X_{t}}(u,v)$ for all $t\ge 0$. We say $v$ is \textit{clockwise} to $u$ and $u$ is \textit{counterclockwise} to $v$ at $t$ if $\delta_{t}(u,v)<n/2$, and $u$ is \textit{opposite} to $v$ if $\delta_{t}(u,v)=n/2$, which can happen only if $n$ is even. Suppose $u$ and $v$ are adjacent in $G$. We say $v$ is a \textit{clockwise neighbor} of $u$ at $t$ if $v$ is clockwise to $u$ at $t$, and $\textit{counterclockwise neighbor}$ at $t$ otherwise. The \textit{width} of $X$ is defined to be the quantity 
\begin{equation*}
	w(X):= \min_{v\in V} \max_{u\in V} \delta_{X}(u,v),
\end{equation*}
which is the length of the shortest path on $\mathbb{Z}_{n}$ that covers all states of the vertices in the configuration. Let $B$ be a subgraph of $G$. We denote by $w_{B}(X)$ the width of the restricted configuration $X|_{V(B)}$ on $B$.
\end{customdef}

\begin{description}[noitemsep]
	\item{(iv)} (\textit{locally monotone coupling}) Two coupled oscillators synchronize regardless of the initial configuration, in such a way that the width is a non-increasing function in time. 
\end{description}

We say that a discrete system of PCOs with this property is \textit{locally monotone}. It turns out that this local monotonicity turns out to be a very important property to establish one of our fundamental observation. Namely, let $\tau$ be a discrete system of PCOs. Note that the dynamic $(G,\tau_{G},X_{0})$ synchronizes if and only if the width $w(X_{t})$ converges to $0$. Now if $\tau$ has locally monotone coupling, then the network on arbitrary connected graph synchronizes once the width becomes small enough $(<n/2)$. This also tells us that synchrony in a locally monotone discrete system of PCOs is stable under small perturbation once established. 

\begin{customlemma}{2.2}[width lemma]\label{widthlemma}
	Let $\tau$ be a locally monotone discrete system of PCOs, $G=(V,E)$ be a graph with a subgraph $B$, and let $X_{0}$ a $n$-configuration on $G$. Suppose $w_{B}(X_{0})<n/2$ and no vertex of $G-B$ pulls any vertex of $B$ on time interval $[0,a]$. Then $w_{B}(X_{t})$ is non-increasing on $[0,a]$. Furthermore, the network $(G,\tau_{G}, X_{0})$ synchronizes regardless of the structure of $G$ if $w(X_{0})<n/2$. 
\end{customlemma}

\qquad In fact, similar results for continuous-time models were observed by different authors. For instance a similar observation was used in \cite{klinglmayr2012guaranteeing} to show synchronization is guaranteed in an invariant subset. On the other hand, a general theorem in \cite{nishimura2011robust} asserts that a similar result applies for systems of continuous-time PCOs where the coupling is of certain type, which includes the type of coupling in our model. Namely, if the initial conditions of the system in \cite{nishimura2011robust} are restricted to be $\phi_i(0)\in \{0,1/n, 2/n, ... , (n-1)/n  \}$, time delay is set to zero and the phase response curve(PRC) is set to $f(x) = -1/n$ for $x\le 0.5$ and $f(x)= 0$ otherwise, the system exactly recreates the behavior between two $n$-state oscillators (the difference of whether the blinking state is $n/2$ or $0$ is entirely superficial) in our model. However, the model in the paper \cite{nishimura2011robust} does not reproduce the exactly same behavior of our discrete model for more than two oscillators, since in their model the coupling effect is \textit{additive} while it is \textit{binary} in ours. Nevertheless, our Lemma \ref{widthlemma} follows from a nearly identical proof to their theorem. So we only give a brief illustration of the key idea in the case of the firefly networks in Figure \ref{width}. 

\qquad Before we get to Figure \ref{width}, we introduce a geometric way of representing discrete systems of PCOs. Usually one visualizes a system of coupled $n$-periodic oscillators with a circular representation, where each oscillator is represented as a revolving dot on a regular $n$-gon. However, it turns out that concentrating on \textit{relative states} rather than on the actual states of the oscillators can be useful. More precisely, if the position of an oscillator $v$ is given by a function $X_{t}(v)$ in time $t$, then its relative position is given by the function $X_{t}(v)-t$ modulo $n$. This can be understood as considering the relative position of each oscillator with respect to an imaginary isolated oscillator, revolving on $\mathbb{Z}_{n}$ regularly without any interruption. The following is a reformulation of $n$-periodic firefly networks in terms of relative states. 

\begin{customdef}{2.3}[relative circular representation] 
Let $G=(V,E)$ be a graph. Let $\alpha$ be an additional singleton vertex, called the \textit{activator}. A map $Y:V\cup \{\alpha\}\rightarrow \mathbb{Z}_{n}$ is called a \textit{relative $n$-configuration} on $G$. Let $S_{\mathtt{rel}}(G,n)$ be the set of all relative $n$ configurations. We say a vertex $v\in V$ is \textit{blinking} if $Y(v)=Y(\alpha)$ where $Y$ is the current relative configuration. The \textit{relative firefly transition map} $\mathtt{t}_{G} : Y\mapsto Y'$ is defined by 
\begin{equation*}
	\mathtt{t}_{G}(Y)(x)=\begin{cases}
Y(x)-1 & \text{if $x=\alpha$}\\
Y(x) -1 & \text{if $x\in V(G)$ has a blinking neighbor $u$ such that $\delta_{Y}(x,u)\le n/2$}\\
Y(x) & \text{otherwise}
\end{cases}
\end{equation*}
The \textit{firefly network} $\mathtt{rel}(G,Y_{0})$ of period $n$ is the discrete-time dynamical system defined on the relative $n$-configuration space $S_{\mathtt{rel}}(G,n)$ by the relative firefly transition map $\mathtt{t}_{G}$. Suppose an initial configuration $Y_{0}$ is given. Then we denote $Y_{t+1}=\mathtt{t}_{G}(Y_{t})$ for $t\ge 0$. 
 \end{customdef}
 
Note that in the above definition, we have used counterclockwise displacement between two oscillators in relative $n$-configuration, which is defined similarly as for non-relative configurations(Definition 2.1). 

\begin{figure*}[h]
     \centering
          \includegraphics[width=0.5 \linewidth]{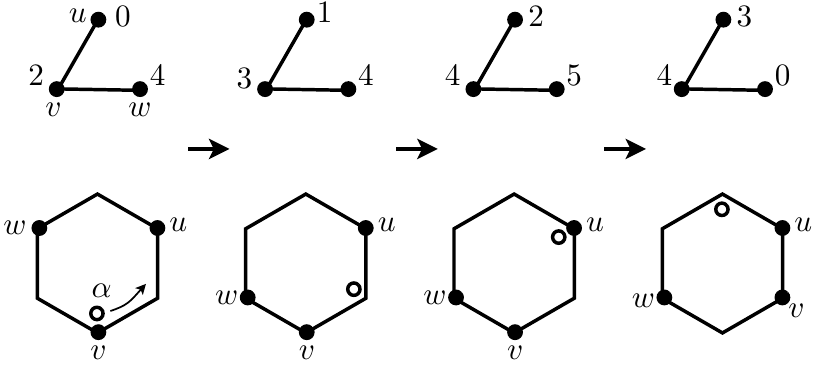}
          \caption{A comparison between the standard and relative circular representation for $n=6$, where the relative circular representation uses hexagon. The latter lacks adjacency information between oscillators. For example, in the third configuration, $u$ only pulls $v$ since it is not adjacent to $w$. If $u$ were adjacent to $w$, then it would have pulled $w$ as well.}
          \label{relative}
\end{figure*}

\begin{figure*}[h]
	\centering
	\includegraphics[width=0.95 \linewidth]{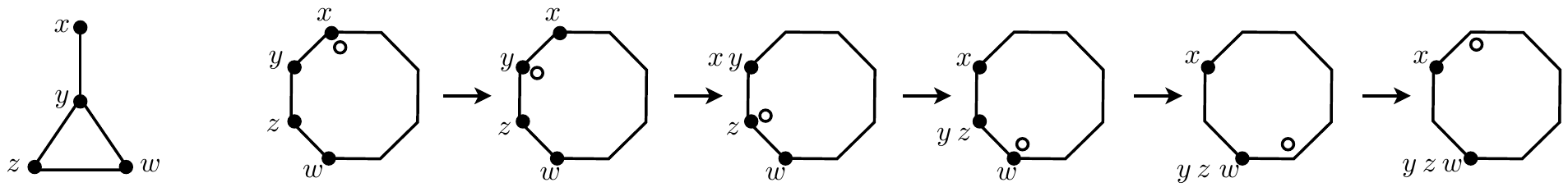}
\caption{An example illustrating the proof of width lemma when $n=8$ and $G=B$. Because of the small width condition on $B$, the head vertices including $x$ does not affect the tail vertices including $w$, and all vertices tend to move toward $w$. The finite size and connection of $G$ will then give the assertion. It is important to note that $x$ never pulls $y$. 
}
	\label{width}
\end{figure*}

 \qquad Using terminologies defined above, we can describe the relative firefly transition map as follows: \textit{at each instant, each vertex moves one step counterclockwise if there is a counterclockwise or opposite blinking neighbor, and does not move otherwise.} Geometrically, each blinking node stretches its left arm on the regular $n$-gon, which is as long as the half of the perimeter, and pulls any of its neighbor within that range. Figure \ref{relative} shows a comparison between the standard and relative circular representation of the firefly network. 
We are going to use those two representations interchangeably. 

\qquad Now we are ready to look at Figure \ref{width}, which conveys the key idea behind Lemma \ref{widthlemma} using the relative circular representation. While the example is for the firefly networks, the same idea was also used in \cite{klinglmayr2012guaranteeing} and \cite{nishimura2011robust} to prove similar results. Namely, if the initial configuration is concentrated on a sufficiently small arc on the unit circle, then the transition map acts as a non-increasing function on the width; it decreases the width to zero if the underlying graph is connected, in which case we have synchrony. For the firefly networks the initial width should be strictly less than the half, and for other similar models it depends on other parameters such as the length of refractory period or maximum delay(see \cite{klinglmayr2012guaranteeing} and \cite{nishimura2011robust}).

\qquad Next, we discuss a characteristic property of our firefly network. Recall it has locally monotone coupling (iv). As in many other models mentioned earlier, a blinking oscillator may either inhibit its hasty(clockwise) neighbors, or excite its lazy(counterclockwise or opposite) neighbors, or both. In our firefly networks, the coupling is \textit{inhibitory} in the following sense:

\begin{description}[noitemsep]
	\item{(v)} (\textit{inhibitory coupling}) A discrete system of PCOs $\tau$ is \textit{inhibitory} if no blinking vertex pulls its counterclockwise neighbor. 
\end{description}

This feature of our model enables inductive arguments on certain class of finite graphs by making a small subgraph irrelevant to the dynamic on the rest. For example, consider the example in Figure \ref{width}. Since the width is always strictly less than half of the perimeter, for the "head" vertex $x$, the only neighbor $y$ is too far to inhibit. So $x$ never pulls $y$, so the dynamic on the triangle with vertices $y,z,w$ is independent of $x$. Hence if we know something about dynamics on the triangle, we can apply that and we deduce some properties of entire dynamic. This notion of restricting global dynamics on subgraphs is given below precisely. 

\begin{customdef}{2.4}
	Let $G=(V,E)$ be a graph and $X_{0}:V\rightarrow C_{n}$ be a $n$-configuration on $G$. Let $H$ be a subgraph of $G$. Let $\tau$ be the firefly transition map as given in Definition 1. We say the dynamic $(G,X_{0})$ \textit{restricts on $H$} if the restriction and transition maps commute, i.e., $\tau_{G}(X_{t})|_{H}=\tau_{H}(X_{t}|_{H})$ for all $t\ge 0$. We say the dynamic $(G,X_{0})$ \textit{restricts on $H$ eventually} if there exists $r\ge 0$ such that $(G,X_{r})$ restricts on $H$. 
\end{customdef}

\section{The branch width lemma, 1-branch pruning, and the path theorem}

\begin{figure*}[h]
     \centering
          \includegraphics[width=0.6 \linewidth]{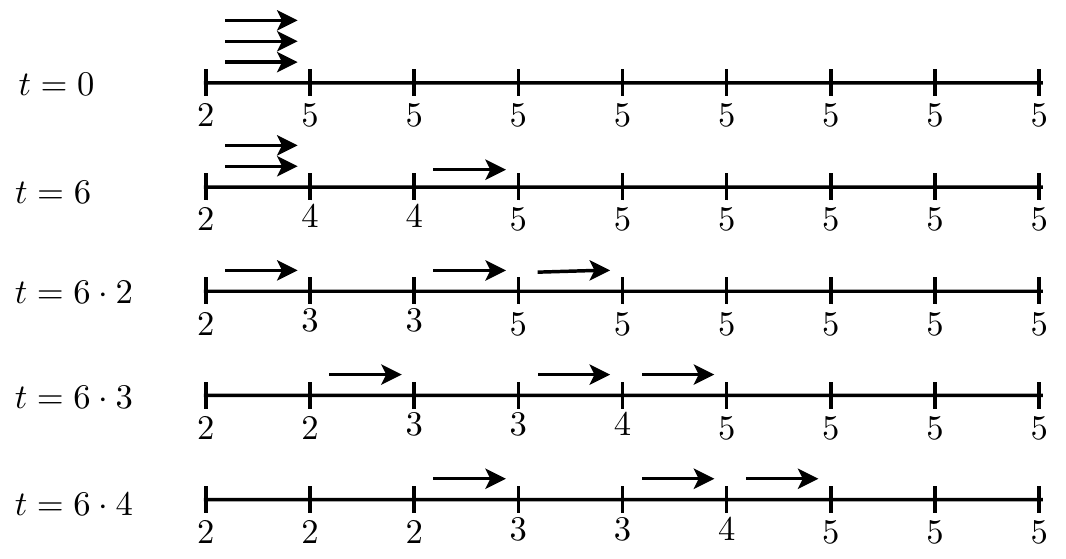}
          \caption{
          	An example of $6$-periodic network on a path of length $m=8$ which takes $6\cdot(6/2-1+8)$ seconds to synchronize. The "color difference" of $n/2$ on the first edge is distributed after $n(n/2-1)$ seconds, and then each "unit difference" moves 1 step right after every $n$ seconds. So we need $nm$ more seconds to take away all particles and achieve synchrony.   
          }
          \label{pathlowerbd}
\end{figure*}

\qquad In this section we establish a key lemma on $n$-periodic firefly networks, which will play a central role in the proof of Theorem \ref{path} and Theorem \ref{blinkingtree}. Our argument for the path theorem goes as follows. Let $P$ be a path of length $m$ and let $n\ge 3$ be period. First note that the lower bound is the time that it takes to synchronize a configuration where all vertices have state $n-1$ but one end vertex, which has the blinking state $b(n)$(See Figure \ref{pathlowerbd}).

For the upper bound, starting with any initial configuration on a finite path, we will observe that the two end vertices of the path do not affect their neighbors eventually, so they become irrelevant to the dynamic of the rest. Hence we can apply inductive arguments on a smaller sub-path. To get this restriction property on paths, we analyze induced local dynamics on the last two vertices of a path, which we call a 1-branch. The general notion of $k$-branch is given below, which will serve a similar role for the tree theorems. 

\begin{customdef}{3.1}
	Let $G=(V,E)$ be a graph. A connected subgraph $S\subseteq G$ is called a \textit{$k$-star} if it has a vertex $v$, called the \textit{center}, such that all the other vertices of $S$ are leaves in $G$. A $k$-star $S$ is called a \text{$k$-branch}, if the center of $S$ has only one neighbor in $G-S$. We may denote a $k$-branch by $B$ rather than by $S$. 
\end{customdef}

Notice that in the definition above, not only $S$ is isomorphic to the canonical star graph, but also the leaves of $S$ must be leaves in $G$. Now the following lemma establishes this restriction property of end vertices of finite paths.  

\begin{customlemma}{3.2}\label{1branch}[1-branch pruning lemma] Let $G=(V,E)$ be a graph with a vertex $w$. Suppose there is a $1$-branch $B$ rooted at $w$ with center $u$ and leaf $v$. Let $n\ge 3$ and let $X_{0}$ be a $n$-configuration on $G$. Let $H=G-v$. Then we have the followings:
\begin{description}[noitemsep]
	\item{(i)} The dynamic $(G,X_{0})$ restricts on $H$ eventually;
	\item{(ii)} Suppose $H$ is $n$-synchronizing. Then $G$ is $n$-synchronizing. Furthermore, if $H$ synchronizes every $n$-configurations in $N_{H}\ge 0$ seconds, then $G$ synchronizes every $n$-configurations in $n^{2}/2+2n-2+N_{H}$ seconds. 
\end{description}
\end{customlemma}

\begin{figure*}[h]
     \centering
          \includegraphics[width=0.22 \linewidth]{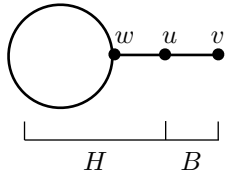}
          \caption{A graph $G$ with a 1-branch $B$}
\end{figure*}

Notice that Theorem \ref{path} follows immediately by an induction, where locally monotone coupling (condition (iv) in Section 2) gives the base case and the above lemma gives the induction step. 

\qquad Before we look into details, we discuss an interesting applications of the theorem. Imagine we have achieved synchrony of oscillators on a graph $G$, and we wish to add a new oscillator $w$ to the already-synchronized network. Notice that this is to consider the dynamic on $G+w$ where all vertices of $G$ have the same state and $w$ may have an arbitrary state. The new dynamic is guaranteed to synchronize by the width lemma(Lemma \ref{widthlemma}) when $n$ is odd, since then  any configuration on $G+w$ with only two states would have width $<n/2$. However, it is possible that the width is exactly $n/2$ when $n$ is even. The following corollary of Theorem \ref{path} implies that we are guaranteed to obtain synchrony in this case too. 

\begin{customcorollary}{3.3}\label{twostates}
     The firefly network $\,\mathtt{rel}(G,Y_{0})$ synchronizes regardless of $G$ if the relative initial configuration consists of only two states.
\end{customcorollary}

\begin{figure*}[h]
     \centering
          \includegraphics[width=11cm,height=2.5cm]{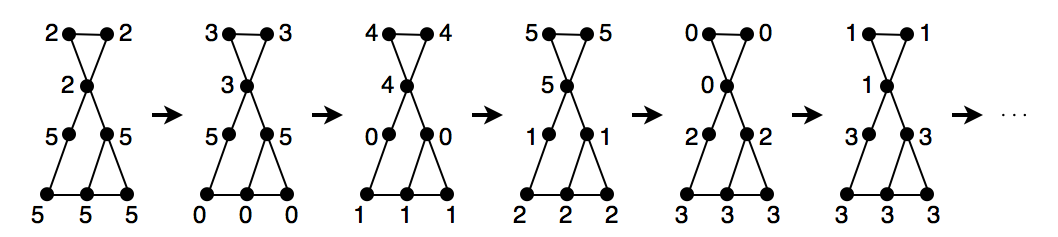}
          \caption{An example of 6-periodic network illustrating Corollary \ref{twostates}. Recall that $b(6)=2$ is the blinking state. In the above example there are only two states 2 and 5 initially. Let $A$ and $B$ be the set of verities with initial state 2 and 5, respectively. Let $A_{0}\subseteq A$ be the neighbors of $B$, and let $A_{1}\subseteq A\setminus A_{0}$ be the neighbors of $A_{0}$, and let $A_{2}\subseteq A\setminus A_{0}\setminus A_{1}$ be the neighbors of $A_{1}$, and so on. We partition $B$ similarly. In the example above, the vertices at the same "level" represents each class $A_{i}$ and $B_{j}$. Then we identify each class into single vertex. This quotient gives a path, and the network on $G$ behaves like that on this path.}
          \label{pathdecomp}
\end{figure*}

\qquad This corollary follows from the following simple but interesting observation; an initial configuration with only two states gives a "path decomposition" of $G$, i.e., a quotient map of $G$ onto a path, and the dynamics on $G$ is essentially that on this path; so it is guaranteed to synchronize by Theorem \ref{path}. This is illustrated in Figure \ref{pathdecomp}.

\qquad The following lemma gives the key observation in this paper. It says that if a graph has a branch, and if the initial configuration has a small width on the branch, then the leaves of the branch become irrelevant to the dynamics on the rest eventually. This is called the \textit{branch width lemma}, which is a variant of the width lemma(Lemma \ref{widthlemma}) using a specific inhibitory structure of the coupling in our model. 

\begin{customlemma}{3.4}[branch width lemma]\label{branchwidthlemma}
     Let $G=(V,E)$ be a graph with a vertex $w$. Suppose there is a $k$-branch $B$ rooted at $w$, with center $v$ and leaves $l_{1},\cdots,l_{k}$, $k\ge 1$. Let $H$ be the graph obtained from $G$ by deleting the leaves of this branch. Let $n\ge 3$ and let $X_{0}$ be a $n$-configuration on $G$. Suppose $w_{B}(X_{0})<n/2-1$. Then we have the followings:
\begin{description}[noitemsep]
	\item{(i)} $v$ is clockwise to all leaves of $B$ at some time $r\le n(w_{B}(X_{0})+1)$ and $w_{B}(X_{r})\le w_{B}(X_{0})$;
	\item{(ii)} $v$ is clockwise to all leaves of $B$ for all $t\ge r$, and $w_{B}(X_{t})\le w_{B}(X_{0})+1$ for all $t\ge 0$;
	\item{(iii)} 	If $v$ is clockwise to all leaves at $t=r$, then the dynamic $(G,X_{r})$ restricts on $H$; 
	\item{(iv)} If $H$ is $n$-synchronizing, then $(G,X_{0})$ synchronizes.
\end{description}
\end{customlemma}

\begin{figure*}[h]
     \centering
          \includegraphics[width=0.22 \linewidth]{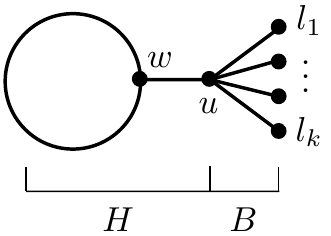}
          \caption{A graph $G$ with a $k$-branch $B$} 
          \label{kbranch}
\end{figure*}         

\begin{figure*}[h]
     \centering
          \includegraphics[width=0.8 \linewidth]{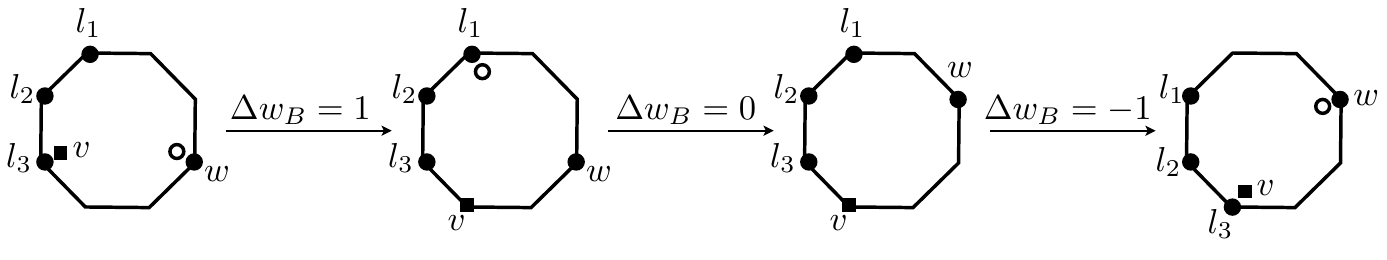}
	\caption{An illustration of branch width recovery for $n=8$ and $k=3$. Once $v$ is at the tail, $w$ can pull $v$ to increase the branch width by 1 but $v$ pulls the head leaf and decrease the branch width by 1, before $w$ blinks again. Note that $w$ can get external pullings from its neighbors different from $v$ but it doesn't affect our argument.}
	\label{branchwidth}
\end{figure*}

\qquad A detailed proof of this lemma is given in the Appendix A, and here we give a quick illustrative explanation. Suppose $n=8$ and $k=3$. Since the coupling is inhibitory, the three leaves of $B$ and the root $w$ only inhibit the center $v$, and during this period, Lemma \ref{widthlemma} (i) keeps the small width on the leaves of $B$. So eventually, we will have a situation as in the first diagram in Figure \ref{branchwidth}, where the branch width $w_{B}:=w|_{B}$ is still strictly less than $n/2-1$ and the center $v$ is at the "tail". Now the root $w$ will pull $v$ occasionally, increasing the branch width by 1. But since we have a wiggle room on the branch width, the increased branch width is still small($<n/2$) and the leaves do not pull $v$ until they blink again. Then the center blinks and pulls all the leaves, decreasing the branch width by 1. Hence the original branch width is recovered, and because of the small width on the branch, the leaves never pull the center. 

\begin{figure*}[h]
     \centering
          \includegraphics[width=0.55 \linewidth]{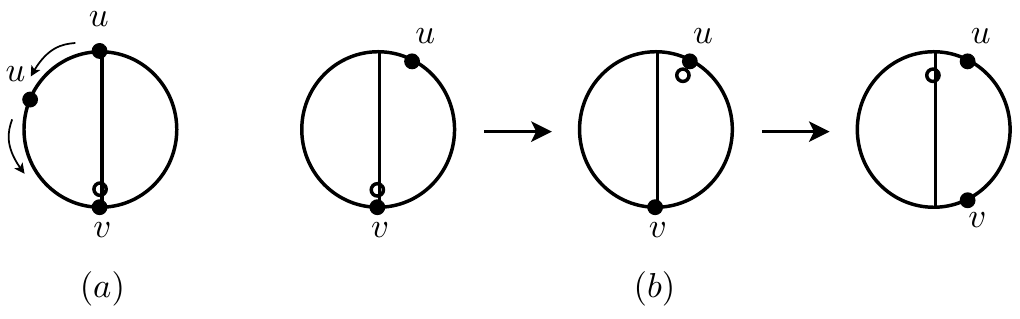}
          \caption{(a) If $\delta_{0}(v,u)\le n/2$ when $u$ first blinks, then $\delta_{t}(v,u)$ monotonically decreases to zero. (b) Otherwise $\delta_{0}(u,v)< n/2$ when $v$ first blinks, and the root $w$ does not pull the center $v$ until the activator moves from $u$ to $v$, and $v$ pulls $u$ to make $\delta_{t}(u,v)<n/2-1$.}
          \label{1pruning}
\end{figure*}

\qquad Note that the 1-brach pruning(Lemma \ref{1branch}) follows easily from the above lemma. Indeed, let $G,w,v$, and $u$ be as in Lemma \ref{1branch}. It suffices to show that eventually, the width on this 1-branch, which is just the minimum of $\delta_{t}(u,v)$ and $\delta_{t}(v,u)$, becomes less than the threshold $n/2-1$. This is illustrated in Figure \ref{1pruning}, and the upper bound in Lemma \ref{1branch} (ii) can be computed easily from the figure. 

\section{Local configurations, Poincar\'e return map, and tree theorems}

\qquad In this section we discuss firefly networks on trees. More specifically, we prove Theorem \ref{treedegree} and Theorem \ref{blinkingtree} for $n\in\{3,4,5\}$. The bottom line of our proofs is the following. Let $T$ be a tree and $X_{0}$ be a $n$-configuration for some $n\in \{3,4,5\}$, and suppose for contrary that $(T,X_{0})$ is a minimal counterexample to Theorem \ref{blinkingtree}. Then an analysis of the induced local dynamic on a particular branch of $T$ would yield that the global dynamic must restrict onto a smaller subtree, which contradicts to the minimality. 

\qquad An important concept for such local analysis is Poincar\'e return map, which is to take "snapshots" of system configurations where a fixed special agent of the system has a particular state, and look into the induced dynamic on the set of such snapshots. This technique effectively reduces dimensionality of system configurations, and was used by Mirollo and Strogatz \cite{mirollo1990synchronization} by taking snapshots only when a particular oscillator blinks. In this paper, we incorporate similar technique to analyze induced local dynamic on a branch, where the center of a fixed branch is taken to be the special vertex. 

\begin{customdef}{4.1}
	Let $G=(V,E)$ be graph, $n\ge 3$, and let $Y$ be a relative $n$-configuration on $G$. Let $v$ be a vertex, $N\subset N_{G}(v)$ be a subset of all neighbors of $v$, and let $N_{v}:=N\cup \{v\}$. Then the local configuration of $X$ on $N_{v}$ is the restriction $Y|_{N_{v}}$. If $N_{v}=N_{G}(v)$, then we write $Y^{v}:=Y|_{N_{v}}$. Fix a relative initial $n$-configuration $Y_{0}$. We say that with respect to the dynamic $\mathtt{rel}(T,Y_{0})$, a local configuration on $N_{v}$ is \textit{recurrent} if it occurs infinitely often, and \textit{transient} otherwise. 
\end{customdef}

The following observation quickly gives us some transient local configurations at some vertex $v$ with a leaf. 

\begin{customlemma}{4.2}[opposite leaf lemma]\label{oppleaf}
	Let $G=(V,E)$ be a graph with a leaf $u$ and its neighbor $v$. Let $Y_{0}$ be any relative initial configuration on $G$. Suppose $Y^{v}$ is any relative local configuration at $v$ where $v$ blinks and the counterclockwise displacement $\delta_{t}(u,v)$ is $\lfloor n/2 \rfloor$. Then $Y^{v}$ is transient. 
\end{customlemma}

\begin{proof}
	This proposition can be best understood by looking at the following cases when $n=6$ in Figure \ref{opposite}. Suppose the first local configuration in Figure \ref{opposite} (a) is recurrent at the 1-branch $B:=v+u$ for some leaf $u$ of $v$. We may back-track for 3 seconds. During this period the center $v$ does not blink so the leaf $u$ is not pulled, and $v$ couldn't have been pulled by any of its neighbors for the first 2 backward iterations, since it was counterclockwise to the activator during that period. Hence the local configuration before three seconds must have been the fourth configuration in Figure \ref{opposite} (a), but the first iteration from this conflicts to the pulling of $u$ on $v$. Thus the shaded local configuration in Figure \ref{opposite} (a) is transient. Since this holds for all 1-branch centered at $v$, the assertion for the $n=\text{even}$ case follows. Figure \ref{opposite} (b) for $n=5$ illustrates the similar argument for the odd period case. 	
\begin{figure*}[h]
	\centering
	\includegraphics[width=0.8 \linewidth]{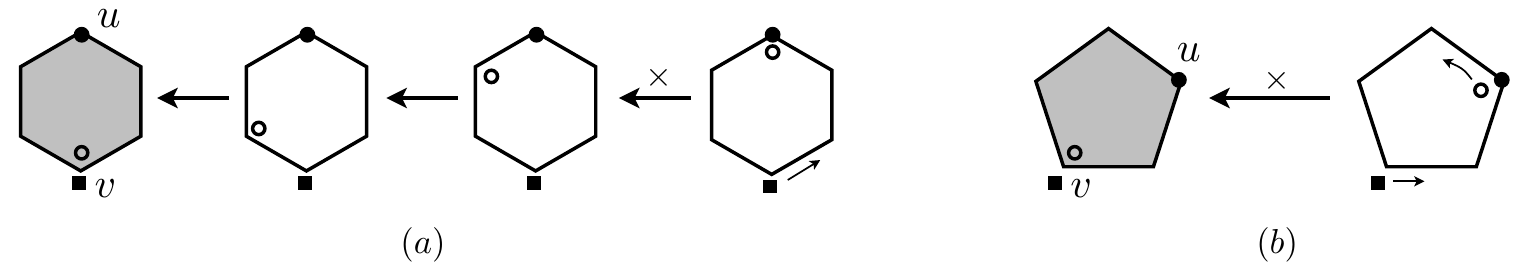}
	\caption{An illustration of the opposite leaf lemma for (a) $n=6$  and (b) $n=5$.  The square and dot represents the relative states of the center $v$ and leaf $u$, respectively. The two shaded local configurations are transient.}
	\label{opposite}
\end{figure*}	
\end{proof}

We call a local configuration $Y|_{B}$ on a branch $B$ \textit{opposite} if $Y|_{B'}$ is such a local configuration in the above proposition for some $B'=v+u$, where $v$ is the center of $B$ and $u$ is some leaf of $B$; we call $Y|_{B}$ \textit{non-opposite} otherwise.

\qquad We start proving Theorem \ref{blinkingtree} for $n=3$. Here we only need Lemma \ref{1branch} and Lemma \ref{oppleaf} for local analysis. 

\begin{proof}[Proof of Theorem \ref{blinkingtree} for $n=3$]
	The "only if" part is trivial. Let us show the "if" part. Let $T=(V,E)$ be a tree, and let $X_{0}:V\rightarrow \mathbb{Z}_{3}$ be a 3-configuration such that every vertex of $T$ blinks in the dynamic $(T,X_{0})$. We wish to show that $(T,X_{0})$ synchronizes. We use an induction on $|V|$. Since $K_{2}$ is 3-synchronizing, we may assume $|V|\ge 3$. If $T$ has a 1-branch, then we are done by the induction hypothesis and Lemma \ref{1branch}. So we may assume there is no 1-branch. Since we can identify any two leaves of the same state with common neighbor, we may assume all leaves of each $k$-star in $T$ have distinct states all time. 
	
\qquad Now let $u$ be a leaf in $T$, and let $v$ be the neighbor of $u$. Since $|V|\ge 3$ and $T-u$ is connected, $v$ has a neighbor $w$ in $T-u$. Let $B$ be a $k$-branch consisting of $v$ and all of its leaf neighbors. Note that $k\ge 2$ by our assumption. By the hypothesis the center $v$ blinks infinitely often, so we can apply Lemma \ref{oppleaf}, which says in this case that eventually, whenever $v$ blinks(has state 1), all leaves of $v$ must have either state 1 or 2. Thus we may assume that $v$ has exactly two leaves which eventually have state 1 and 2 whenever $v$ blinks. But at any such instant, $v$ pulls the leaf of state 2 and synchronizes its two leaves. This contradicts our assumption that no two leaves of $v$ ever have the same state. This shows the assertion. 
 \end{proof}

\qquad To proceed more concisely in proving Theorem \ref{blinkingtree} for $n=4$ or $5$, we consider the following class of configurations for which our inductive argument may not work. 

\begin{customdef}{4.3}
	Let $G=(V,E)$ be a graph. A $n$-configuration $X_{0}:V\rightarrow \mathbb{Z}_{n}$ on $G$ is \textit{irreducible} if the dynamic $(G,X)$ never restricts to a proper subgraph of $G$ of at least $2$ vertices. If $X_{0}$ is irreducible, we say $(G,X_{0})$ is an \textit{irreducible dynamic}. 
\end{customdef}

Notice that if $(G,X_{0})$ synchronizes, then $X_{0}$ cannot be irreducible, since after the synchrony the dynamic can be restricted on any proper subgraph of $G$. The next proposition tells us that any initial configuration for a minimal counterexample for Theorem \ref{blinkingtree} is irreducible if every vertex blinks infinitely often in the dynamic. 

\begin{customprop}{4.4}
	Let $T=(V,E)$ be a tree, and suppose that there exists a $n$-configuration $X_{0}:V\rightarrow \mathbb{Z}_{n}$ such that every vertex blinks infinitely often in $(T,X_{0})$ but the network does not synchronize. Further assume that $T$ is a smallest such tree. Then $X_{0}$ is irreducible. 
\end{customprop}

\begin{proof}
	Suppose not. Then there exists a proper subtree $T'$ of $T$ with $\ge 2$ vertices and an integer $r\ge 0$ such that the network $(T,X_{r})$ restricts on $T'$. Then every vertex of $T'$ blinks infinitely often in $(T',X_{r}|_{T'})$ since they do so in the larger network $(T,X_{0})$. Then by the minimality of $T$ the restriction synchronizes. After the synchrony on $T'$, we may identify the vertices of $T'$. In other words, eventually, we can contract $T'$ to a single vertex without affecting the dynamic. Denote the resulting graph by $T/T'$, which is a proper minor of $T'$. So $T/T'$ is a tree which is strictly smaller than $T$. Clearly every vertex of $T/T'$ blinks infinitely often in the induced dynamic, so we get a synchrony on $T/T'$. But this tells us $(T,X_{0})$ eventually reaches synchrony, which is a contradiction. Therefore $X_{0}$ must be irreducible. 
\end{proof}

\qquad Hence to obtain Theorem \ref{blinkingtree} for period $4$ and $5$, it suffices to show that if there is a minimal counterexample $(T,X_{0})$ then $X_{0}$ cannot be irreducible. In the proof of $n=3$ case, we used the fact that a certain local configuration on a branch is transient, and then Lemma \ref{1branch} to restrict the global dynamic on a smaller subtree. 

\begin{customprop} {4.5}
	Let $T=(V,E)$ be a tree, $n\ge 3$, and let $X_{0}$ be a $n$-configuration which is not irreducible. Suppose $T$ has a $k$-star $S$ and let $v$ be its center. Then any local configuration $Y|_{S}$ is transient if it satisfies either of the following conditions:
\begin{description}[noitemsep]	
	\item{(i)} Some two distinct leaves of $B$ have the same state in $Y$;
	\item{(ii)} $Y|_{S}$ is opposite.
\end{description}
Furthermore, if $S=B$ is a branch, then $Y|_{B}$ is also transient if it satisfies either of the following conditions:
\begin{description}[noitemsep]	
	\item{(iii)} Only a single state is used for the leaves of $B$;
	\item{(iv)} $Y|_{B}$ has width $<n/2-1$.
\end{description}
\end{customprop}

\begin{proof}
	For (i), suppose that two leaves $x,y$ of $B$ have the same state at some point. Then they will always have the same state later on, so we may restrict the global dynamic onto $T-x$, for example. This contradicts the irreducibility. Lemma \ref{oppleaf} shows (ii) is transient. Now suppose $S=B$ is a branch. Then the 1-branch pruning lemma(Lemma \ref{1branch}) shows (iii) is transient. Finally, (iv) follows from the branch width lemma(Lemma \ref{branchwidthlemma}). 
\end{proof}

\begin{customprop}{4.6}\label{treerecurrentprop}
	Let $G=(V,E)$ be a graph with an induced $k$-star $S$ centered at a vertex $v$. Let $n\in \{4,5\}$, and suppose $X_{0}$ is an irreducible $n$-configuration on $G$ such that $v$ blinks infinitely often in the dynamic.  Then we have the followings:
\begin{description}[noitemsep]
	\item{(i)} If $n=4$, then all local configurations on $S$ where $w$ blinks are transient except Figure \ref{treerecurrent} $a$;
	\item{(ii)} If $n=5$ and if $S=B$ is a branch with root $w$, then all local configurations on $B$ where $v$ blinks are transient except Figure \ref{treerecurrent} $b$.
\end{description}

\begin{figure*}[h]
     \centering
          \includegraphics[width=0.3 \linewidth]{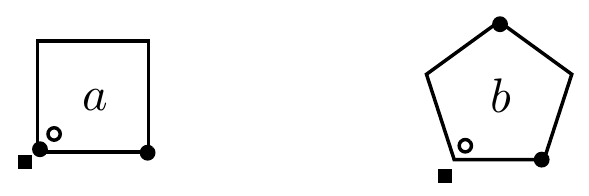}
          \caption{Possibly recurrent local configurations on stars for $n=4$ and on branches for $n=5$.}
          \label{treerecurrent}
\end{figure*}   

\end{customprop}

\begin{proof}
	For $n=4$ it follows easily from Proposition 4.8 (i) and (ii). We give a detailed argument for $n=5$ case. Suppose $S$ is a branch $B$ rooted at $w$. So the neighbors of $v$ are the leaves of $B$ and the root $w$. By the irreducibility, we may assume that no two leaves ever have the same state. Consider the local configurations on $S$ where $v$ blinks. First of all, there are 15 non-opposite local configurations on $S$ as in Figure \ref{5tree}.

\begin{figure*}[h]
     \centering
          \includegraphics[width = 0.95 \linewidth]{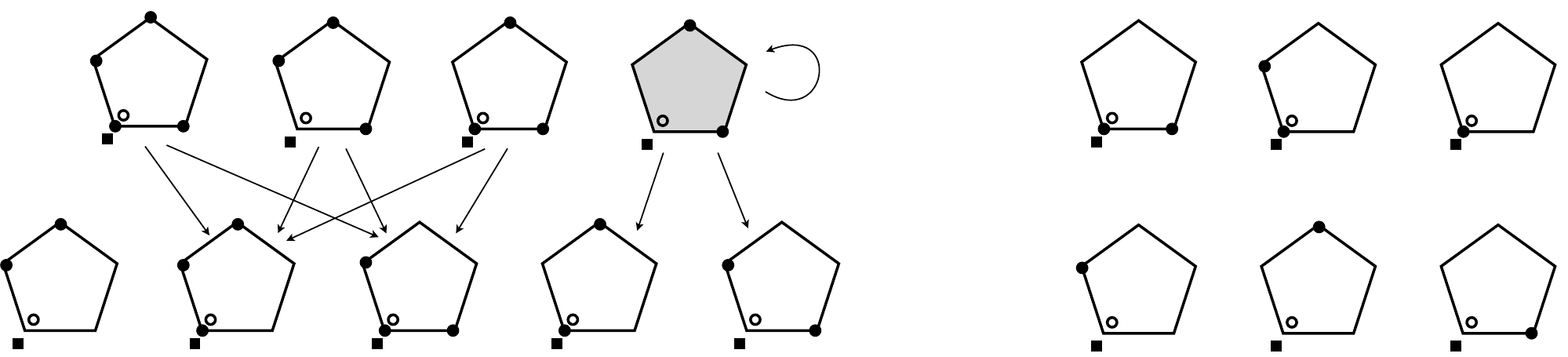}
          \caption{Non-opposite local configurations on a branch $B$ with some of the possible transitions between them. The dots represent leaves, and the square represent the center of the branch. If the configuration is irreducible, only the shaded one is recurrent.}
          \label{5tree}
\end{figure*}

Notice that the 6 local configurations on the right are transient by (iii) and (iv) of the previous proposition. Furthermore, the first five local configurations in the second row lead to some of the 6 local configurations on the right, after the center $v$ pulls some of its leaves. Hence they are also transient.

\qquad Now to rule out the first three in the first row, we consider all possible transitions from them. Remember that $v$ could be pulled by the root $w$ during this transition, so the next position of $v$ is uncertain in the local sense. But note that after the center $v$ pulls the leaves ahead of itself, there is no further change on the leaves until $v$ blinks again, so the next position of leaves are determined locally. Hence the first three in the first row can only lead to the second and third in the second row, which are transient. But since $v$ blinks infinitely often, at least one local configuration where $v $ blinks must be recurrent. Thus the assertion follows. 
\end{proof}

Now Theorem \ref{blinkingtree} for $n=5$ follows directly from the following lemma:

\begin{customlemma}{4.7}[branch pruning for $n=5$]\label{branchpruning5}
     Let $G=(V,E)$ be a graph with a vertex $w$. Suppose there is a $k$-branch $B$ rooted at $w$ with center $v$(see Figure \ref{kbranch}). Then there is no irreducible $5$-configuration on $G$ such that every vertex blinks infinitely often in the dynamic. 
\end{customlemma}

\begin{proof}[Sketch of proof]
	Let $X_{0}$ be any 5-configuration on $G$ where all vertices blink infinitely often in the dynamic. It suffices to show that the dynamic on $G$ restricts on $G-B$ eventually. For this we show that the center $v$ does not pull $w$ eventually. By Proposition \ref{treerecurrentprop}, $B$ has two leaves which have states 1 and 4 whenever $v$ blinks eventually. This specific dynamic on $B$ is possible only if $w$ blinks at some specific moments, and an easy analysis shows that this yields $w$ must have state 0 whenever $v$ blinks(a similar analysis is given for the $n=4$ case). Hence $w$ is not pulled by $v$ eventually, as desired. 
\end{proof}

\qquad It now remains to show Theorem \ref{blinkingtree} for $n=4$. Interestingly, there is no 4-periodic counterpart of Lemma \ref{branchpruning5} That is, not every branch can be pruned out under the assumption of irreducible configuration where all vertices blink infinitely often. However, this pruning is possible when the graph is a tree and the branch is at the end of a longest path. 

\begin{customlemma}{4.8}\label{4treelemma}
		Let $G=(V,E)$ be a graph. Suppose that $G$ has an irreducible $4$-configuration $X_{0}$ such that every vertex of $G$ blinks infinitely often in $(G,X_{0})$. Suppose $G$ has a $k$-star $S$ for $k\ge 2$ with center $v$. Then the followings are true:
\begin{description}[noitemsep]
	\item{(i)} $k=2$ and whenever $v$ blinks(has state $1$), its two leaves have state 0 and 1. 
	\item{(ii)} If $S=B$ is a branch in $G$, then in the limit cycle the local dynamic on $B$ and its root repeats the following sequence:
	\begin{equation}
\begin{matrix}
\text{center | leaves }	&	\textbf{1}|01 & 2|12 & 2|23 & 3|30 & 3|01 & 3|12 & 3|23 & 0|30 & \textbf{1}|01 \\
\text{root\qquad\qquad}	&3 &  3	&   0 & \textbf{1}  & 2 & * & * & * & 3 
\end{matrix}
\end{equation}
\end{description}	
	\item{(iii)} In case of (ii), the root of $B$ must have at least three neighbors in $G-B$. 
	\item{(iv)} There are no two branches rooted at the same vertex. 
\end{customlemma}

\begin{proof}

(i) directly follows from Proposition 4.9. Now suppose $S$ is a branch with root, say, $w$. For the other parts, we analyze the actual transition between the unique recurrent local configuration $a$ in Figure \ref{treerecurrent}. The analysis in Figure \ref{4treeex} shows that starting from this local configuration, the center $v$ must be pulled by the root $w$ at the third or fourth iteration, and each transition takes exactly 8 seconds. 
\begin{figure*}[h]
     \centering
          \includegraphics[width=0.7 \linewidth]{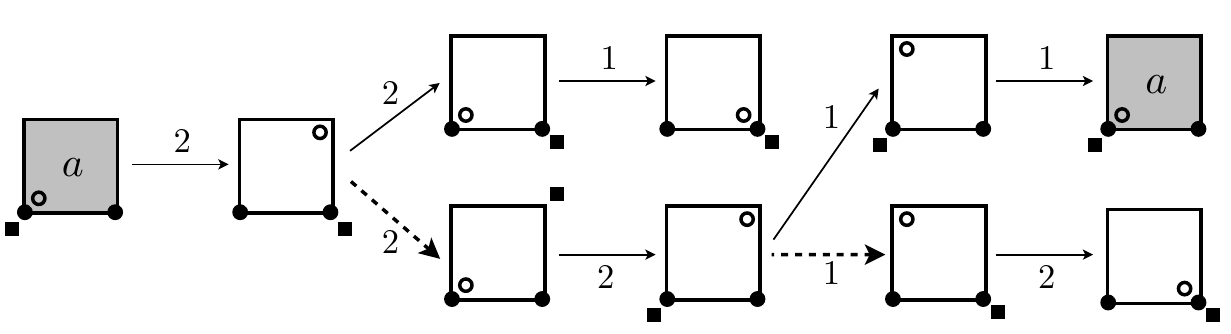}
          \caption{ Actual transition between the unique recurrent local configuration $a$ in Figure \ref{treerecurrent} when $S=B$ is a branch. The center is pulled by one of its leaves in 2 seconds, and for the next 2 seconds it may or may not be pulled by the root $w$, so it splits into two cases. The upper one is for no pulling from $w$, which ends up with a wrong local configuration. Hence the center $v$ must be pulled by $w$ at third or fourth iterations. On the other hand, the dashed arrow to the third column indicates a transition where $v$ is pulled by $w$. Since $w$ blinks at most once in every 4 seconds, $v$ can be pulled by $w$ at most once in every 4 seconds. Hence there is no external pulling in the following 2 seconds. Then the case splits into two as before, and this time $w$ should not pull the center to end up with the correct local configuration. The whole transition takes exactly 8 seconds. 
          }
          \label{4treeex}
\end{figure*}

Note that In the standard representation, the recurrent local configuration $a$ in Figure \ref{treerecurrent} can be represented as $1|01$, by which we mean $\text{state of center}\,|\, \text{state of leaves}$. Hence the above analysis shows that $1|01$ must lead to $3|01$ after 4 seconds. We consider the first 4 iterations starting from $1|01$ in standard representation:
\begin{equation}
\begin{matrix}
\text{center | leaves }	&	\textbf{1}|01 & 2|12 & *|23 & *|30 & 3|01 \\
\text{root\qquad \qquad}	&x_{1} &  x_{2}	&   x_{3} & x_{4}  & x_{5}
\end{matrix}
\end{equation}
where $1\in \{x_{3},x_{4}\}$. However, $x_{3}=1$ yields $x_{2}=0$ and $x_{1}=3$ which is impossible since the blinking center then must have been pulled $x_{1}=3$ so that $x_{2}=3$. Therefore $x_{4}=1$, and consequently, $x_{1}=3$. Thus the 8 iterations and each configurations on the branch and root is given by (9). This shows (ii). 

\qquad For (iii), observe that in the sequence (9), the root $w$ must be pulled 3 times in the last four seconds. Since $v$ do not pull $w$ during this iterations, and since $w$ can get at most one pulling from each of its neighbor in every 4 seconds, we conclude that $w$ must have at least 3 neighbors different from $v$. This shows (iii). 

\qquad To show (iv), suppose there are two branches $B_{1}$ and $B_{2}$ rooted at $w$. By (ii), the dynamics on those branches in the limit cycle are given by (9). Note that none of $*$'s in (9) is $1$ for trivial reason. Now whenever $w$ blinks, the centers and leaves in both branches must have the same configuration, namely, $3|30$. This determines the dynamics on those branches completely, and we see that they must have the identical dynamic. Thus the dynamic on $G$ reduces to $G-B_{1}$, which contradicts the irreducibility. This shows (iv). 
\end{proof}

\begin{proof}[Proof of Theorem \ref{blinkingtree} for $n=4$]
	It suffices to show the "if" part. Let $(T,X_{0})$ be a minimal counterexample to Theorem \ref{blinkingtree} for $n=4$. It suffices to show that $X_{0}$ is not irreducible. Suppose the contrary. We may assume $|V|\ge 3$. We first show that $T$ must have an "inevitable" local structure as in Figure \ref{4treeproof}. Let $P$ be a \textit{longest} path in $T$. It has length at least $2$ since $|V|\ge 3$. Let $u$ be an end of $P$, $v$ the neighbor of $u$ in $P$, and $w$ the neighbor of $v$ in $P-u$. Then by the choice of $v$ and Lemma \ref{4treelemma} (i), $v$ and its leaves form a 2-branch rooted at $w$. Let us denote this branch by $B$. By Lemma \ref{4treelemma}, the dynamic on $B$ and the root $w$ repeats the sequence (9). By Lemma \ref{4treelemma} (iv), there is no other branch rooted at $w$. It then follows from our choice of $w$ from a maximal path $P$, that every neighbor of $w$ not in $P$ is a leaf. We conclude that $w$ must have at least two leaves by Lemma \ref{4treelemma} (iii). Then by Lemma \ref{4treelemma} (i), $w$ and its leaves form a 2-star.

\begin{figure*}[h]
     \centering
          \includegraphics[width=0.26 \linewidth]{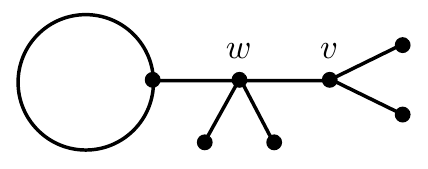}
          \caption{ A local structure of $T$ assuming $(T,X_{0})$ is a minimal counterexample for Theorem \ref{blinkingtree} for $n=4$
          }
          \label{4treeproof}
\end{figure*}

\qquad Now by Lemma \ref{4treelemma} (i), eventually, the two leaves of $w$ must have state 0 and 1 whenever $v$ blinks. Now we insert the states of leaves of $w$ into the sequence (9). The first four iterations would then be as follows:
\begin{equation}
\begin{matrix}
\text{$u$ | leaves of $u$}	&	\textbf{1}|01 & 2|12 & 2|23 & 3|30 & 3|01 \\
\text{$w$ | leaves of $w$}	&3|\textbf{1}2 &  3|23&   0|30 & \textbf{1}|01  &2|12  
\end{matrix}
\end{equation}
Recall that the above sequence describes four iterations starting from an arbitrary instant when $v$ blinks. Notice that there is a blinking leaf of $w$ at the first instant. Thus we conclude that whenever $v$ blinks $w$ has a blinking neighbor so that the pulling of $v$ on $w$ is redundant. Thus the dynamic on $T$ restricts on $T-B$ eventually, contrary to our assumption. This shows the assertion. 
\end{proof}

\qquad We have seen in the introduction that Theorem \ref{treedegree} follows directly from Theorem \ref{blinkingtree} and Lemma \ref{deglemma}. The proof of Lemma \ref{deglemma} is independent from our storyline, so we give it in Appendix B.

\section{Randomized self-stabilization of firefly network on arbitrary connected graphs} 

\qquad Let $\tau$ be a $n$-periodic discrete system of PCOs for some $n\ge 3$. In the case of firefly networks, we have seen that non-synchronizing limit cycles depend heavily on the symmetry of the graph and configurations on it. In order to break this symmetry to enhance synchronization, we introduce randomness to our GCA model in the following way. For a given connected graph $G=(V,E)$, consider an edge weighting $\epsilon:E\rightarrow (0,1)$. Now each edge $e$ in $G$ will be present with a fixed probability $\epsilon(e)$ independently at each instance in the dynamic $(G,\tau_{G})$, so two adjacent oscillators along the edge $e$ will now see each other with the associated probability. In other words, we are introducing stochastic reception of pulses. This introduces randomness to our deterministic GCA model and turns it into a Markov chain, which we denote by a triple $(G,\epsilon,\tau_{G})$. We identify all constant configurations into a single state called \textit{sync}, and take 
\begin{equation*}
\mathcal{S}:=\{X:V\rightarrow \mathbb{Z}_{n}\,|\, \text{$X$ is non-constant} \}\cup \{\textit{sync}\} 
\end{equation*}
as the state space of our Markov chain. Now clearly \textit{sync} is an absorbing state, meaning that once the system state is \textit{sync}, then it is so thereafter. In fact, \textit{sync} is the only absorbing state in this Markov chain. Given that, Theorem \ref{stochastic} follows from an elementary argument in Markov chain theory. 

\qquad Showing that \textit{sync} is the unique absorbing state is based on the following simple observation. Note that it suffices to show that for any given initial configuration $X$, there is a positive probability $p_{X}>0$ such that the Markov chain enters $\textit{sync}$ eventually. To this end, suppose we have achieved synchrony on a connected subgraph $H\subset G$. By the connection of $G$, we can pick a vertex $w$ in $G-H$ that is adjacent to some vertex of $H$. Since the states on $H $ has been synchronized, there are only two states on $H+w$. So if all edges in $H+w$ are present and the bridging edges between $H+w$ and the rest are absent for a sufficient amount of time, $H+w$ behaves as a deterministic network isolated from the rest, so we have synchrony on it by Corollary \ref{twostates}. We repeat this process until we expand a partial synchrony to the entire $G$. Hence one can synchronize arbitrary configuration $X$ with some positive probability through this process. Note that the time that it takes for each step is bounded above by the upper bound given in Theorem \ref{path}. Then an elementary calculation gives an upper bound on the expected time until absorption. However, the upper bound that we get from this argument does appear to be far from optimal.

\qquad The above argument works for any discrete system of PCOs for which the two-state corollary(Corollary \ref{twostates}) holds. While a general locally monotone discrete system of PCOs may lack this property, it is easy to modify this argument to work for arbitrary locally monotone discrete system of PCOs, by using Lemma \ref{widthlemma} instead of Corollary \ref{twostates}, which holds for any locally monotone discrete system of PCOs. Moreover, we may give the stochasticity not by assigning a fixed potability to be absent for each edge, but for each vertex, as in \cite{klinglmayr2012guaranteeing}. A similar argument easily applies, so we obtain the following generalization of Theorem \ref{stochastic}:

\begin{customthm}{5.1}\label{random}
	Let $\tau$ be a $n$-periodic locally monotone discrete system of PCOs for any $n\ge 3$ and let $G=(V,E)$ be a connected graph. We give stochasticity by assuming either of the followings:
	\begin{description}[noitemsep]
		\item{(i)} (stochastic emission) Each vertex $v\in V$ is present independently at each instant with a fixed probability $p_{v}\in (0,1)$;
		\item{(ii)} (stochastic reception) Each edge $e\in E$ is present independently at each instant with a fixed probability $p_{e}\in (0,1)$.
	\end{description}
Then the Markov chain $(G,\tau_{G})$ is an absorbing chain with sync being the unique absorbing state. In particular, every $n$-configuration on $G$ synchronizes with probability 1 in finite expected time. 
\end{customthm}

\section{Concluding remarks}

\qquad We defined a GCA model for pulse-coupled oscillators and studied their network behavior, mainly focused on various conditions for synchrony. Taking the advantage that the dynamic of individual oscillators is extremely simple, we were able to obtained conditions on initial configurations and network topologies that guarantee synchronization. Paths were generic to our model in the sense that every finite path is $n$-synchronizing for all $n\ge 3$. We then studied to what extent this self-stabilization property on paths extends, and we obtained a local-global principle on tree networks for period $n\le 6$(a proof for $n=6$ case was omitted). We also showed that any $n$-periodic firefly network on random networks synchronizes with high probability starting from an arbitrary initial configuration, where random networks are given by introducing independent random errors either to the vertices or edges of connected graphs. 

\qquad The remainder of this section contains a description of directions of future research. Our first cornerstone was Theorem \ref{path}. An obvious extension of this result is to consider infinite path, namely, the integer lattice $\mathbb{Z}$, instead of finite paths. From the example in Figure \ref{pathlowerbd}, however, it is apparent that one can have a non-synchronizing initial configuration on $\mathbb{Z}$ for any period $n$; e.g., a "wave of pullings" could propagate from $-\infty$ to $\infty$. An appropriate type of question to ask in this case might be as follows: starting from a uniform product measure on $\mathbb{Z}$, and fixing a finite interval $I$, does one have synchronization on $I$ with probability 1? If so, how would such a probability scale in time?

\qquad For low periods $n\le 6$, we have seen that the firefly transition map synchronizes every initial configuration on an arbitrary fixed tree, given that the maximum degree is less than the period. However, we have seen that Theorem \ref{blinkingtree} is not valid for $n=7$ in Figure \ref{n7counterex}, and verifying the theorem for $n\ge 8$ is open. On the other hand, we have also seen that cycles and cliques are not synchronizing in general. Indeed, Dolev \cite{dolev2000self} discusses that given any distributed synchronization algorithm, one can always find a non-synchronizing configuration on some cycle. Hence, one might think that not containing cycle as a subgraph is a critical factor for a network to be synchronizing. However, $K_{3}$ is 6-synchronizing and the "shovel graph", obtained by vertex-summing $K_{3}$ at the end of a path, is also 6-synchronizing. Thus just having a cycle in the network does not necessarily mean that the network is not 6-synchronizing. It is a future goal to obtain a complete characterization of $n$-synchronizing graphs. Expanding the class of $n$-synchronizing graphs is of special interest in the view point of self-stabilizing networks, since it would allow us to design a fault-tolerant and self-synchronizing system with variety of network topologies.

\qquad In section 5 we incorporated a certain stochasticity into our deterministic model and obtained absorbing Markov chains with synchrony being the unique absorbing state. Having established this universal synchrony with high probability, it is then interesting to ask what is the expected time until synchrony. While for each graph $G$ we can get exact expected time until absorption using an elementary Markov chain theory, the recursive argument using Corollary \ref{twostates} or Lemma \ref{widthlemma} gives a trivial upper bound for the expected time until synchrony, which only depends on the number of vertices, maximum degree of $G$, and the period. However, this trivial upper bound is too crude to be precisely stated in this paper. On the other hand, take $G=K_{m}$ and assume stochastic reception with $p_{e}\equiv p(n)\in (0,1)$. Then the resulting Markov chain is a firefly network on a sequence of Erd\'os-Renyi random graphs $G(m,p)$. It would be interesting to study phase transitions in this model with respect to the order of $p$. 

\qquad Finally, one can also study different transition maps, in comparison with our firefly transition map. For any class of connected graph $\mathcal{H}$, call a transition map $\tau$ on the space of $n$-configurations $\mathcal{H}$-type if it synchronizes every $n$-configuration on all $H\in \mathcal{H}$. Our firefly transition map, for example, is a path-type which synchronizes tree networks for low periods given a degree condition. One can then study $\mathcal{H}$-type transition maps for various choices of $\mathcal{H}$. Recall that, somewhat contrary to our intuition, the all-to-all networks were not synchronizing in general; $K_{4}$ is not 6-synchronizing as we have seen in Figure \ref{ex1} (d) in the introduction, and the $2n+1$-configuration using states $0,n$, and $2n$ on $K_{3}$ does not synchronize for all $n\ge 2$. So any clique-type transition map, if any, must be essentially different from our firefly transition map. However, notice that by Dolev's argument $\mathcal{H}$ cannot contain the class of cycles. Studying different types of transition maps should be useful for various application of different nature. \\

\appendix
\renewcommand*{\appendixname}{}
\numberwithin{equation}{section}
\numberwithin{figure}{section}
\numberwithin{table}{section}

\setcounter{MaxMatrixCols}{20}

\section{Proof of Lemma \ref{branchwidthlemma}}

\begin{proof}[Proof of Lemma \ref{branchwidthlemma}]
	Observe that if any two leaves have the same state at some point, then they will always have the same state since they get same input from $u$ and they have no other neighbors. Hence we may identify any two leaves after they have the same state. Suppose $X_{0}$ is an $n$-configuration on $G$ with $w_{B}(X_{0})<n/2-1$. Call a vertex of $B$ \textit{at the head}(\textit{tail}) at time $t$ if it is counterclockwise(clockwise) to all the other vertices of $B$ at time $t$. 
	
\qquad We first show (i). If $u$ is at the tail, clearly we may take $r=0$. Otherwise, there exists some leaf at the tail, say $l_{k}$ without loss of generality, which is strictly clockwise to $u$. Observe that both the root $w$ and any leaf strictly clockwise to $u$, if any, will pull $u$ towards the tail. So there will be some $r\ge 0$ such that $u$ is at the tail at time $r$. We may take $r$ as small as possible. On the time interval $[0,r)$, Lemma \ref{widthlemma} tells us that $w_{B}(X_{t})$ is non-increasing as long as $w$ does not affect $u$, but when $w$ does affect $u$, $w_{B}(X_{t})$ would not increase since $l_{k}$ is strictly clockwise to $u$. Thus we have $w_{B}(X_{r})\le w_{B}(X_{0})$. Furthermore, we have the upper bound $r\le n(w_{B}(X_{0})+1)$ since it takes $\le n$ seconds for $l_{k}$ to blink for the first time and it needs to pull $u$ by $\le w_{B}(X_{0})$ steps, and $l_{k}$ blinks once in every $n$ seconds on time interval $[0,r]$. This shows (i). Without loss of generality, let $l_{1}$ be leaf at the head at time $t=r$(see Figure \ref{branchwidth}).

\qquad Next we show (ii). By part (i), It suffices to show that for all $t\ge r$, $u$ is clockwise to all leaves of $B$ and $w_{B}(X_{t})\le w_{B}(X_{r})+1$. The idea is that if $w$ pulls $u$ to increase the branch width by 1 after time $r$, then $u$ will pull the head leaf $l_{1}$ before it gets pulled by $w$ once more and recovers the original branch width(see Figure 7). If $w$ never pulls $u$ for $t\ge r$, then the assertion follows immediately. So suppose at some time $t\ge r$ that $w$ blinks and $u$ is counterclockwise to $w$. We are going to see what will happen until $w$ blinks again. $u$ will be pulled toward $w$, increasing the branch width by 1, but which is $\le w_{B}(X_{0})+1\le n/2$. So the branch width is non-increasing until $w$ blinks again. In fact, $u$ will blink before $w$ does so, and it will pull all the strictly clockwise leaves, in particular, $l_{1}$, and decrease the branch width by 1. Note the branch has recovered its original branch width and $u$ is still at the tail. The assertion is clear if $w$ never blinks again, and otherwise we are back to the previous case. This shows (ii).
	
\qquad By the hypothesis and (ii), we have $w_{B}(X_{t})<n/2$ for all $t\ge 0$ and $u$ is at the tail for all $t\ge r$. Hence no leaves of $B$ pulls $u$ on time interval $[r,\infty)$, so the dynamic $(G,X_{r})$ restricts on $H$. If $H$ is $n$-synchronizing, this means eventually, all vertices of $H$ will have the same state as $u$. Then the branch width becomes the total width, and we have synchrony on $G$ by Lemma \ref{widthlemma}. This shows (iii) and (iv). 
\end{proof}

\section{Proof of Lemma \ref{deglemma}}

\begin{proof}[Proof of Lemma \ref{deglemma}]
	Suppose for contrary that $u$ does not blink after some $t_{0}\in \mathbb{N}$. Consider the clockwise displacement $\delta_{t}(u,\alpha)$ from $u$ to the activator $\alpha$. Note that by definition, $u$ blinks at $t$ if and only if $\delta_{t}(u,\alpha)=0$. Since $u$ does not blink after $t=t_{0}$, the displacement is positive for $t>t_{0}$. Note that the displacement is non-increasing until the activator catches up to $u$, and it strictly decreases whenever no neighbor of $u$ is blinking, since then $u$ does not move while $\alpha$ moves toward $u$. Hence we may assume at $t= t_{1}\ge t_{0}$, the displacement stabilizes to its minimum and at least one neighbor of $u$ blinks at each second $t\ge t_{1}$. 
	
\qquad Since the degree of $u$ is at most $n-1$, there is always an unoccupied state on $\mathbb{Z}_{n}$ by the neighbors of $u$. Suppose there is an empty spot with counterclockwise displacement $\le n/2$ from the activator as in Figure \ref{degree} (a). Then no neighbor of $u$ whose state is between the empty spot and the activator moves while the activator proceeds toward the empty spot, so when the activator occupies the empty spot, there is no blinking neighbor of $u$, contrary to our assumption. Hence each state with counterclockwise displacement $\le n/2$ from the activator must be occupied by at least one neighbor of $u$ for all $t\ge t_{1}$. Now let $h(t)$ be the number of neighbors of $u$ that are strictly behind the activator at time $t+t_{1}$, that is, that are with clockwise displacement $<n/2$ from the activator. Note that any oscillators that were blinking at $t=t_{1}$ fall behind the activator at $t=t_{1}+1$, so $h(1)\ge 1$(see Figure \ref{degree} (b)). Note also that this neighbor strictly behind the activator will be counted by the function $h$ in the next $b(n)-1$ seconds. As the activator proceeds one more step counterclockwise, at least one more oscillator falls behind the activator and we have $h(2)\ge 2$. Suppose $n=2m$. The similar argument gives $h(m-1)\ge m-1$. But by the assumption all the $m+1$ spots within counterclockwise displacement $\le n/2$ from the activator must be occupied by the neighbors of $u$. It follows that at $t=t_{1}+m-1$, there are at least $2m=n$ neighbors of $u$, a contradiction. The similar argument leads to a contradiction for the case $n=2m+1$. Thus $u$ must blink infinitely often. 
\begin{figure*}[h]
     \centering
          \includegraphics[width = 0.8 \linewidth]{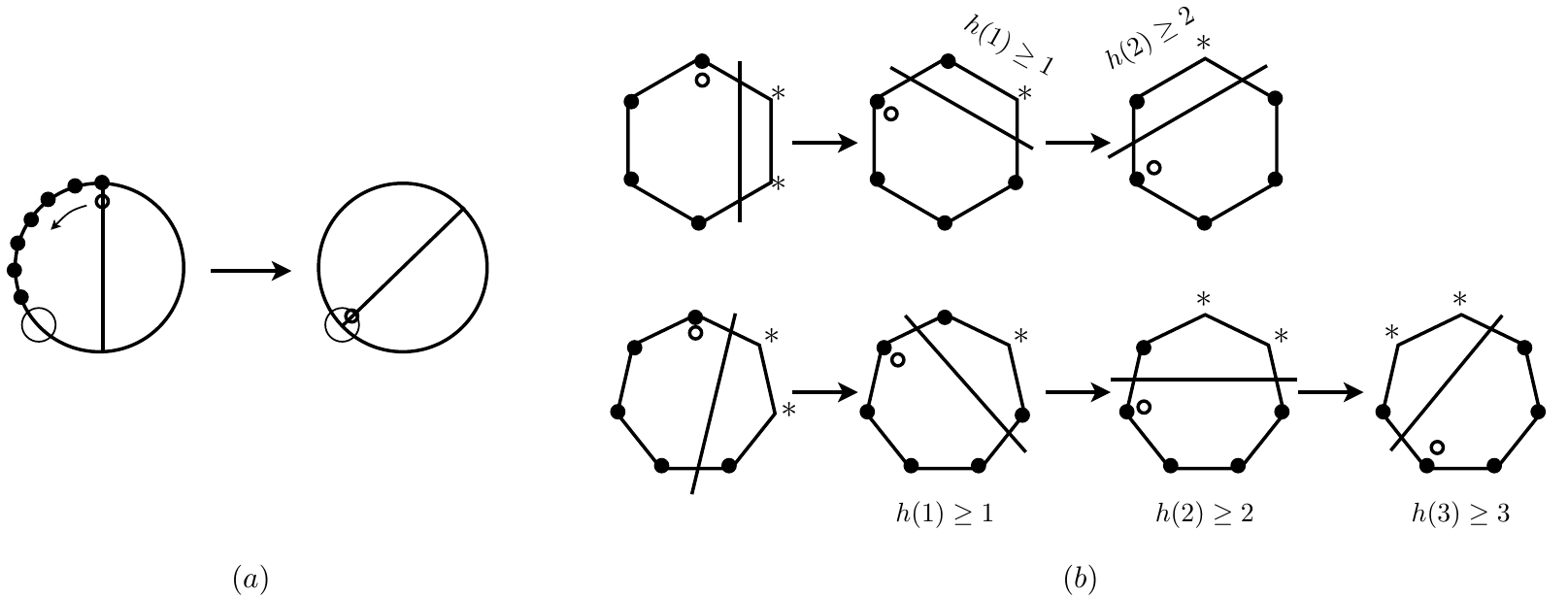}
          \caption{(a) The activator occupies any forward empty spot without pulling any neighbor of $u$ to the spot. (b) More oscillators fall behind the activator but the states counterclockwise to the activator must be all occupied. $*$ indicates the spots that may or may not be occupied by neighbors of $u$.}
          \label{degree}
\end{figure*}         

\end{proof}

 \section*{Acknowledgement}
 
We give special thanks to David Sivakoff for his priceless advices, and also to Woong Kook and Hyuk Kim for their encouragement on this work. In addition, we appreciate the referees for their insightful comments and suggestions.

%% The Appendices part is started with the command \appendix;
%% appendix sections are then done as normal sections
%% \appendix

%% \section{}
%% \label{}

%% If you have bibdatabase file and want bibtex to generate the
%% bibitems, please use
%%

%\section*{References}

\bibliographystyle{elsarticle-harv}   % this means that the order of references
			    % is dtermined by the order in which the
			    % \cite and \nocite commands appear

\bibliography{mybib}  % list here all the bibliographies that
			     % you need. 

%% else use the following coding to input the bibitems directly in the
%% TeX file.

%\begin{thebibliography}{00}

%% \bibitem[Author(year)]{label}
%% Text of bibliographic item

%\bibitem[ ()]{}
%\end{thebibliography}

%%%%%%%%%%%%figures

\end{document}